\documentclass{ws-ijmpa}

\begin{document}

\markboth{KORAY KARACA, SEL\c{C}UK BAYIN}
{An Open Singularity-Free Cosmological Model with Inflation}

\catchline{}{}{}

\title{AN OPEN SINGULARITY-FREE COSMOLOGICAL MODEL WITH INFLATION}

\author{\footnotesize KORAY KARACA\footnote{
korayk@newton.physics.metu.edu.tr}}

\address{Physics Department, Middle East Technical University, \\
Ankara, 06531, Turkey}

\author{SEL\c{C}UK BAYIN\footnote{bayin@metu.edu.tr}}
\address{Physics Department, Middle East Technical University, \\
Ankara, 06531, Turkey}

\maketitle

\pub{Received (27 June 2002)}{}
\begin{abstract}
~~~~In the light of recent observations which point to an open universe  
$(\Omega_{0}<1)$, we construct an open singularity-free cosmological model
by
reconsidering
a
model originally constructed for a closed universe. Our model starts from
a nonsingular state called prematter, governed by an inflationary equation
of state
$P=(\gamma_{p}-1)\rho$ where $\gamma_{p}$ $(\simeq 10^{-3})$ is a small
positive parameter
representing the initial vacuum dominance of the universe. Unlike the    
closed models universe cannot be initially static hence, starts with an   
initial expansion rate represented by the initial value of the Hubble
constant H(0). Therefore, our model is a two-parameter universe model
$(\gamma_{p},H(0))$. Comparing the predictions of this model for the 
present properties of the universe with the recent observational results,
we argue that the model constructed in this work could be used as a
realistic universe model.

We started the evolution of a flat universe from a nonsingular state
called prematter which is governed by an inflationary
equation of state $P=(\gamma_{p}-1)\rho$ where $\gamma_{p}$ represents the
initial vacuum dominance of the universe. The evolution of the
universe-except in the prematter era-is
affected neither by the initial vacuum dominance nor the initial expansion
rate of the universe. On the other hand, present properties of the
universe such as Hubble constant, age and density are sensitive to the
temperature at the decoupling 

 Over a range between
value between 50 and 80 $Km\cdot sec^{-1}\cdot Mpc^{-1}$ for the present
value of the Hubble constant $(H_{0})$. Assuming that the thermal history
of the universe is independent from its geometry, the above range could be
considered as transition range for the decoupling temperature  

\end{abstract}

\section{Introduction}	

Although the standard cosmological model has been succesful in explaining
the homogenous expansion of the universe and the 2.7 K cosmic microwave
background radiation, it has some shortcomings like the initial
singularity, horizon (or causality), flatness, homogeneity and isotropy
problems.\cite{Guth} Among these problems the initial singularity
problem may  
be the one which weakens the model much more than the others, in the sense
that it causes infinities in the physical quantities such as the density,
pressure and temperature. Since these infinities cannot be accepted by any
model which claims to be physical, during the past two decades, several
authors have considered the possibility of describing the universe with a
singularity-free cosmological
model. \cite{Rosen,Israelit,Starkovich,Bayin}

Recent observations point out that the universe has an $\Omega
_{0}  (\Omega \equiv \rho /\rho _{c}$ , $\rho _{c}\equiv 3H^{2}/8\pi )$
value
lower than unity which means that the universe is spatially
open. \cite{Henry,Eke,Bahcall,Bartelmann,Cooray} In the light of these, we
propose an open
singularity-free
cosmological model with the same motivations as in Ref. 5. However,
initially  
static condition $\left( \dot{a}(0)=0\right) $ used by the closed universe
models can no longer be used in the open universe case since it leads to
negative energy densities. For this reason in our model universe starts
expanding with
an initial velocity. Hence, we have a two-parameter universe model in
which
one of the parameters characterizes the equation of state used to describe
the prematter era, and the other one corresponds to the initial expansion
rate of the universe.

\section{Description of the Model}
\subsection{Field Equations}

Our model describes an open, spatially homogenous and isotropic universe
with a space-time geometry given by the Robertson-Walker (RW) line
element:

\begin{equation}
ds^{2}=dt^{2}-a^{2}(t)\left. \left[
{\displaystyle{dr^{2} \over 1+r^{2}}}%
+r^{2}\left( d\theta ^{2}+\sin ^{2}\theta d\phi
^{2}\right) \right. \right] ,
\end{equation}
where $(t,r,\theta ,\phi )$ are the comoving coordinates, $a(t)$ is the
scale factor which represents the size of the universe. \ \

For the RW metric, Einstein field equations lead us to the following
differential
equations:
\begin{equation}
\left(
{\displaystyle{\dot{a} \over a}}%
\right) ^{2}-%
{\displaystyle{1 \over a^{2}}}%
=%
{\displaystyle{8\pi  \over 3}}%
\rho ,
\end{equation}

\begin{equation}
2%
{\displaystyle{\ddot{a} \over a}}%
+\left(
{\displaystyle{\dot{a} \over a}}%
\right) ^{2}-%
{\displaystyle{1 \over a^{2}}}%
=-8\pi P,
\end{equation}
where $\rho $ and $P$ are the energy density and pressure in the universe 
respectively,
where we have used perfect fluid energy momentum tensor in comoving
coordinates. Here a dot denotes differentiation with\ respect to the
cosmic
time $t$ and we use the units so that $M_{pl}=G^{-1/2}=1,$ where $M_{pl}$
is
the Planck mass.
Combining Eqs. (2) and (3) and using an equation of state given as
\begin{equation}
P=\left( \gamma -1\right) \rho ,
\end{equation}
one obtains the following relation:

\begin{equation}
{\displaystyle{\ddot{a} \over a}}%
+4\pi \left( \gamma -\frac{2}{3}\right) \rho =0.
\end{equation}
Where, $\gamma $ is assumed to be a constant parameter during each era in
the history of the universe.

We can eliminate $\rho $ between Eqs. (2) and (5) and obtain an equation
involving only the scale factor $a(t)$:

\begin{equation}
\left(
{\displaystyle{\ddot{a} \over a}}%
\right) +\left(
{\displaystyle{3 \over 2}}%
\gamma -1\right) \left(
{\displaystyle{\dot{a}^{2}-1 \over a^{2}}}%
\right) =0.
\end{equation}
To solve this equation for any $\gamma $, it is advantageous to work in
conformal time $\eta $ which is defined as

\begin{equation}
dt=a\left( \eta \right) d\eta .
\end{equation} 

Using this transformation, Eq. (6) is solved to give  two linearly
independent solutions given as
\begin{equation}
a\left( \eta \right) =a_{0}\left[ \sinh \left( \eta c+\delta
\right) \right]
^{1/c},
\end{equation}
\begin{equation}
a\left( \eta \right) =a_{0}\left[ \cosh \left( \eta c+\delta
\right) \right]
^{1/c},
\end{equation}
where
\begin{equation}
c=%
{\displaystyle{3 \over 2}}%
\gamma -1.
\end{equation}
and $a_{0}$ and $\delta $ are the integration constants.

These solutions correspond to different signs of the energy density $\rho
$.
Assuming that the universe is filled with positive energy, we eliminate
Eq. 
(9) and consider only Eq. (8) as the solution for the scale factor of
the
universe.
In this work, we model the universe as a series of perfect fluid eras
connected by first order phase transitions. According to the
considerations
of this scenario, the universe starts expanding with a period of rapid
expansion
called inflationary era. This period is characterized by an equation of
state in the form $P\simeq -\rho $ and inflation arises due to this
vacuum-like characteristic of the equation of  
state. Since matter would be under extreme conditions during this period
and
behaves very differently from the ordinary matter, it is called
``prematter''. \cite{Israelit} During this period due to the unusual
characteristic
of
the equation of state, temperature increases although the universe expands
enormously. We assume that inflation continues until the temperature
reaches
the maximum allowed temperature i.e. the Planck temperature $%
T_{pl}=1.4169\cdot 10^{32}$ $K$ . This behavior, which does not
necessitate
a ``re-heating mechanism'' as in the other models of the universe, follows
from the vacuum like characteristic of the equation of state used to
describe the universe in this era.

\subsection { Boundary Conditions and the Solutions For the Scale
Factor}

Initially, we assume that the universe is in a vacuum-like state and has a
limiting density called the Planck density $\left( \rho _{pl}\right) ,$
which is first formulated by Markov as a universal law of
nature. \cite{Markov} 
Since our model describes a universe starting from a finite size
and
density, the initial expansion rate must be taken as positive. Hence we
take
the initial expansion rate as

\begin{equation}
a^{\prime }\left( 0\right) =v,
\end{equation}
where $v$ is some positive constant.

Solutions for the scale factor in different eras are:\
\begin{equation}
a(\eta )=\left\{
\begin{array}{ll}
a_{0}^{(p)}[\sinh (c_{p}\eta +\delta _{p})\ ]^{1/c_{p}}\  & \ \ 0\leq \eta
\leq \eta _{r}, \\
a_{0}^{(r)}[\sinh (\eta +\delta _{r})] & \eta _{r}\leq \eta \leq \eta
_{m},
\\
a_{0}^{(m)}[\sinh (\eta /2+\delta _{m})]^{2}\  & \ \eta _{m}\leq \eta .
\end{array}
\right.
\end{equation}  
where \textit{p}, \textit{r}, and \textit{m} denote the prematter,
radiation and matter eras
respectively. We
next impose the boundary condition that the scale factor and its
derivative are continuous at points $\eta _{r}$ and $\eta _{m},$ where the
phase transitions take place. This determines the integration constants
as,

\begin{equation}
a_{0}^{(p)}=\left[
{\displaystyle{\sqrt{v^{2}-a^{2}(0)} \over a(0)}}%
\right] ^{\frac{1}{c_{p}}+1}d^{-1/2},
\end{equation}

\begin{equation}
\delta _{p}=\ln \sqrt{%
{\displaystyle{v+a(0) \over v-a(0)}}%
},
\end{equation}

\begin{equation}
a_{0}^{(r)}=a_{0}^{(p)}[\sinh (c_{p}\eta _{r}+\delta
_{p})]^{\frac{1}{c_{p}}%
-1},
\end{equation}

\begin{equation}
\delta _{r}=\left( c_{p}-1\right) \eta _{r}+\delta _{p},
\end{equation}

\begin{equation}
a_{0}^{\left( m\right) }=\frac{a_{0}^{(r)}}{\sinh \left( \eta _{m}+\delta
_{r}\right) },
\end{equation}

\begin{equation}
\delta _{m}=\eta _{m}/2+\delta _{r}.
\end{equation}

As seen from the solutions for the scale factor, the model that we propose
is a two-parameter model. The parameters are the $c_{p}$ value and the
initial value of the Hubble constant $(H\left( 0\right) )$. The former  
determines the amount of inflation that the universe has experienced
during
the prematter era and the latter is related to the initial expansion rate
of
the universe.

We can find \ the comoving times corresponding to the conformal times
$\eta
_{r},$ $\eta _{m},$ and $\eta _{now}$ by using the definition given in Eq.
(7). Assuming that $t=0$ at $\eta =0$ we get from Eq. (8)

\begin{equation}
t\left( \eta \right) =a_{0}\int_{0}^{\eta }\left[ \sinh \left( \eta
^{\prime
}c+\delta \right) \right] ^{1/c}d\eta ^{\prime }.
\end{equation}
This integral depends on the values of $c$ and has to be computed
numerically in the prematter era. Whereas, for the radiation $\left(
c_{r}=1\right) $ and matter $\left( c_{m}=1/2\right) $ eras,
Eq. (18) could
be integrated to yield analytical expressions. The expressions for the
comoving times corresponding to $\eta _{r},$ $\eta _{m}$ and $\eta _{now}$
are

\begin{equation}
t_{r}=a_{0}^{\left( p\right) }\int_{0}^{\eta _{r}}[\sinh \left( c_{p}\eta
+\delta _{p}\right) ]^{1/c_{p}}d\eta ,
\end{equation}

\begin{equation}
t_{m}=t_{r}+a_{0}^{\left( r\right) }\left[ \cosh \left( \eta _{m}+\delta
_{r}\right) -\cosh \left( \eta _{r}+\delta _{r}\right) \right] ,
\end{equation}

\begin{equation}
t_{now}=t_{m}+%
{\displaystyle{a_{0}^{\left( m\right) } \over 2}}%
\left[ \sinh \left( \eta _{now}+2\delta _{m}\right) -\sinh \left( \eta  
_{m}+2\delta _{m}\right) +\left( \eta _{m}-\eta _{now}\right) \right] .
\end{equation}

Hubble constants at $\eta _{r},\eta _{m}$ and $\eta _{now}$ could now be
obtained by using
\begin{equation}
H(\eta)=\frac{a^{\prime}(\eta)}{a^2(\eta)},
\end{equation}
as
\begin{equation}
H(\eta_{r})=\frac{a^\prime(\eta_{r})}{a^2(\eta_{r})}=
\frac{(9.2503\cdot10^{29})~cosh(c_{p}\eta_{r}+\delta_{p})} 
{a_{0}^{(p)}\left[sinh(c_{p}\eta_{r}+
\delta_{p}\right]^{\frac{1}{c_{p}}+1}}~Km\cdot s^{-1}\cdot Mpc^{-1},
\end{equation}   
\begin{equation}
H(\eta_{m})=\frac{a^\prime(\eta_{m})}{a^2(\eta_{m})}=
\frac{(9.2503\cdot10^{29})~cosh(\eta_{m}+\delta_{r})}
{a_{0}^{(r)}sinh^2(\eta_{m}+\delta_{r})}~Km\cdot s^{-1}\cdot Mpc^{-1},
\end{equation}  
\begin{equation}
H(\eta_{now})=\frac{a^\prime(\eta_{now})}{a^2(\eta_{now})}=
\frac{(9.2503\cdot10^{29})~cosh(\eta_{now}/2+\delta_{m})}
{a_{0}^{(m)}sinh^3(\eta_{now}/2+\delta_{m})}~Km\cdot s^{-1}\cdot Mpc^{-1}.
\end{equation}   

The evolution of the energy density is described by 
\begin{equation}
\rho^{\prime}+3\gamma\frac{a^{\prime}}{a}\rho=0.
\end{equation}  
Eq. (26) could then be solved to yield
\begin{equation}
\frac{\rho(\eta_{f})}{\rho(\eta_{i})}=\left(
\frac{a(\eta_{f})}{a(\eta_{i})}\right)^{3\gamma},
\end{equation}
where $\eta_{i}$ and  $\eta_{f}$ are the initial and final instants of any
conformal time interval in a given era. Choosing
$\eta_{i}, \eta_{f}=(0,\eta_{r}),~ (\eta_{r},\eta_{m}),~
(\eta_{m}, \eta_{now})$ we obtain respectively  
\begin{equation}
\rho(\eta_{r})=\left(\frac{a(0)}{a(\eta_{r})}\right)^{3\gamma_{p}}\rho_{pl},
\end{equation}  
\begin{equation}
\rho(\eta_{m})=\left(\frac{a(\eta_{r})}{a(\eta_{m})}\right)^4\rho(\eta_{r}),
\end{equation}
\begin{equation}
\rho(\eta_{now})=\left(\frac{a(\eta_{m})}{a(\eta_{now})}\right)^3\rho(\eta_{m}).
\end{equation}

\section{Numerical Results}
In order to see this dependence and the development of the   
universe in this model, we changed one of the parameters while fixing the
other and provided some numerical results. This allowed us to identify
the character of the dependency of the model to each parameter.
First, we fix $H(0)$ to $2.3427\cdot 10^{63}$ 
$Km\cdot s^{-1}\cdot Mpc^{-1}$,
and changed $\gamma_{p}$ in such a way that it took values between
$2.0000\cdot 10^{-3}$ and $2.0500\cdot 10^{-3}.$ It is to be noted that
in this broad range, the value of Hubble constant has a wide spectrum
between $49.2304$ $Km\cdot s^{-1}\cdot Mpc^{-1}$ and $96.5196$
$Km\cdot s^{-1}\cdot Mpc^{-1}.$ Next,
we set $\gamma_{p}$ to $2.0350\cdot 10^{-3}$ and varied $H(0)$ in the
range between $2.6941\cdot 10^{63}$ $Km\cdot s^{-1}\cdot Mpc^{-1}$  and
$2.1084\cdot 10^{63}$ $Km\cdot s^{-1}\cdot Mpc^{-1}$. The corresponding
values for the Hubble constant vary between $50.1328$
$Km\cdot s^{-1}\cdot Mpc^{-1}$ and $91.6791$  
$Km\cdot s^{-1}\cdot Mpc^{-1}.$

\section{Conclusion}

We construct an open singularity-free cosmological model with the
assumptions that the universe is initially in a vacuum like state and the
physical quantities are limited by their Planck values. Evolution of the
temperature of the universe is governed by its expansion. During the
prematter era, temperature increases due to inflation and the
universe is characterized by a vacuum like equation of state in the
form  
$%
P=(\gamma _{p}-1)\rho $ where $\gamma _{p}\sim 10^{-3}$ is a
parameter    
which
determines the vacuum dominance of the early universe. Hence, the
model we
construct is a two-parameter universe model, the parameters being
$H(0)$
and $\gamma _{p}.$
. This era ends when   
the temperature reaches the maximum allowed temperature $T_{pl}.$ Then the
universe enters the radiation era and starts cooling down. This cooling 
continues during the matter era described by the standard model.

The so-called flatness problem is not present in this model since
the initial value of $\Omega $ is not set to unity. In the standard model,  
such a precise initial condition has to be assumed without explanation to   
produce a universe resembling the actual one. In this model, the universe
starts its journey with an $\Omega $ value no matter how close to unity
and
during the prematter era $\Omega $ is driven toward unity. At the end of
this era, the universe is nearly flat since its $\Omega $ value is very   
close to one. This fact could be attributed to the inflation mechanism
which
causes the universe to expand enormously in a very small time interval
(Planck time) and thus become flatter than at the beginning. During the
radiation and matter eras, $\Omega $ is driven away unity which apparently
displays the open character of the space-time geometry of the model.

In this work, we have not been interested in the microphysics of
the inflationary era. This includes quantum mechanical investigation of
the
material content of the early universe and the form of the scalar field 
potential responsible for the inflation. However, with this simple form,
the
scenario proposed in this work might provide a guidance for the future
more 
complete versions of the open inflation.

\end{document}